\pdfoutput=1
\documentclass[11pt, a4paper]{article}

\usepackage{jcappub}

\usepackage{comment}
\usepackage{amsmath}
\usepackage{amsfonts}
\usepackage{amssymb}
\usepackage{graphicx}
\usepackage{mathrsfs}
\usepackage{latexsym}
\usepackage{color}
\usepackage{slashed, cancel}
\usepackage{hyperref}
\usepackage{cleveref}
\usepackage{float}
\usepackage{tikz}
\usepackage{bbold}
\usepackage{soul} 
\usepackage{mathtools}
\usepackage{support-caption} 
\usepackage{enumitem}
\usepackage{journals}
\allowdisplaybreaks
\usepackage{setspace}

\usepackage[utf8]{inputenc} 
\usepackage[version=3]{mhchem}
\hypersetup{urlcolor=blue, colorlinks=true} 

\usepackage{fancyhdr}
\usepackage[caption=false]{subfig}
\graphicspath{{figs/}}
\usepackage{dcolumn}
\usepackage{bm}
\usepackage{parskip}
\usepackage{booktabs}
\usepackage{multirow}

\numberwithin{equation}{section}

\definecolor{rossos}{rgb}{0.8,0.2,0.3}
\definecolor{bluscuro}{rgb}{0.15, 0.2, .85}
\definecolor{bluchiaro}{cmyk}{1,.3,0.,0.1}
\hypersetup{colorlinks, citecolor=bluscuro, linkcolor=bluscuro, urlcolor=bluscuro}

\makeatother   
\newcommand{\GeV}{{\rm \,GeV}}

\newcommand{\keV}{{\rm \,keV}}

\newcommand{\cm}{{\rm \,cm}}

 \def\be   {\begin{equation}}   \def\ee   {\end{equation}}
 \def\ba   {\begin{array}}      \def\ea   {\end{array}}
 \def\bea  {\begin{eqnarray}}   \def\eea  {\end{eqnarray}}
 \def\bean {\begin{eqnarray*}}  \def\eean {\end{eqnarray*}}

\begin{document}

\today

\title{Inelastic Dark Matter and the SABRE Experiment}

\author[a,1]{Madeleine J. Zurowski,\note{Corresponding author, \tt madeleine.zurowski@unimelb.edu.au}}
\author[a,2]{Elisabetta Barberio,\note{\tt barberio@unimelb.edu.au}}
\author[b,3]{Giorgio Busoni \note{ \tt giorgio.busoni@mpi-hd.mpg.de}}


\affiliation[a]{ARC Centre of Excellence for Dark Matter Particle Physics \\
School of Physics, The University of Melbourne, Victoria 3010, Australia}
\affiliation[b]{Max-Planck-Institut fur Kernphysik, Saupfercheckweg 1, 69117 Heidelberg, Germany.}


\abstract{We present here the sensitivity of the SABRE (Sodium iodide with Active Background REjection) experiment to benchmark proton-philic, spin dependent, Inelastic Dark Matter models previously proposed due to their lowered tension with existing experimental results. We perform fits to cross section, mass, and mass splitting values to find the best fit to DAMA/LIBRA data for these models. In this analysis, we consider the Standard Halo Model (SHM), as well as an interesting extension upon it, the SHM+Stream distribution, to investigate the influence of the Dark Matter velocity distribution upon experimental sensitivity and whether or not its consideration may be able to help relieve the present experimental tension. Based on our analysis, SABRE should be sensitive to all the three benchmark models within 3-5 years of data taking.}

\maketitle
\section{Introduction}
\label{sec:level1}

The presence of Dark Matter (DM), some non-luminous, non-baryonic material within the Universe, is well supported by astrophysical observations dating back to the 1930s \cite{Zwicky:1933gu}. While in principle this additional matter could be either of Astrophysical or Particle origin, observations such as \cite{Allsman:2000kg,Tisserand:2006zx} tend to favour the latter case. Based on these, this particle or particles must be a new addition to the Standard Model of Particle Physics (SM). Among the many possible particle candidates, Weakly Interactive Massive Particles (WIMPs) have drawn a lot of attention, due to the possibility of naturally accounting for the right abundance through thermal production in the early universe - the so called ``WIMP Miracle" \cite{Gondolo:1990dk,Bertone:2004pz}. There are three main ways to try to detect WIMP candidates: Direct Detection (DD), which tries to observe the scattering of a DM particle with SM matter, Indirect Detection (ID), which aims to detect the products of annihilations of DM particles, and Collider Searches, where one infers the presence of long-lived particles in the products of high energy collisions taking place in hadron or electron colliders. 

Although the existence of DM is largely accepted within the physics community, to date all these search methods lack experimental observation of it. The only notable exceptions to this are the results published by the DAMA Collaboration over the last 15 years. For nearly two decades DAMA have consistently observed a modulating signal consistent with a WIMP DM presence in the galaxy with a combined significance of $12.9\sigma$, using a NaI(Tl) target \cite{Bernabei2018}. To date, no other collaboration has manged to replicate or observe this modulation \cite{Patrignani:2016xqp}, a point of great tension within DM physics.  However, as DAMA is a DD experiment, it relies on a process that is highly dependent on the relative masses of the DM and SM target, the velocity of the incoming DM, and the process or particle that mediates the interaction. Typical DD targets are noble gases or crystal scintillators that produce a detectable signal when a collision with DM occurs, and a DM interaction with one target does not guarantee interaction with another due to their different nucleon composition. Thus, in order to conclusively refute or support DAMA's results, tests must be conducted in a model independent way using an experiment with the same target and method, as non-NaI targets must assume some model a priori for comparative analysis \cite{Kahlhoefer:2018knc}.


What remains compelling about the DAMA results despite the ongoing tension is their modulating nature. The presence of DM within the galaxy is expected to produce a signature modulation due to the Earth's rotation around the Sun as it moves through the galactic DM. When the motion of the Earth opposes the solar velocity, the relative velocity between the DM and the target will be at a minimum, compared to when it is moving in the same direction as the Sun. This effect will occur regardless of the actual particle physics interaction taking place, as it is dependent only on the velocity distribution of DM, not any particular target. Although the modulating component is typically at least an order of magnitude smaller than the average, because most background contributions are constant in time a modulating signal can be easier to observe. The only signals that contribute to a modulating background are potential seasonal effects - particles or processes, such as cosmic muons, that change with the seasons, and so will have the same period as a DM modulation. However, unlike DM (which is galactic in origin), seasonal modulations will have a different peak depending on which hemisphere measurements are taken in, while DM should produce identical signals. These background modulations can be modelled or measured separately to exclude them from the DM data set. The modulation that DAMA has observed above background has a phase, amplitude, and period consistent with galactic DM.
Thus, to verify DAMA's claimed signal, another NaI-based detector needs to observe the same modulation. At present, the candidate experiments that are equipped to perform this model independent analysis are SABRE \cite{SABREpop}, COSINE-100 \cite{COSINE2019} and ANAIS-112 \cite{Amare2019}.

If a NaI based experiment does manage to observe the DAMA modulation signal, the null results of other experiments still require an explanation. 
The absence of any observation in agreement points towards more exotic models for DM, as the DAMA result is excluded for the most simplistic case - an elastic, spin independent interaction \cite{Baum:2018ekm}. One such model that has received increased attention of late is that of Inelastic DM (IDM) \cite{ArkaniHamed:2008qn,Cui:2009xq,Chang:2008gd}, where the DM particles scatter inelastically off of nucleons into higher mass states. In this case, the assumption is made that DM is constructed from at least two distinct but related particles, $\chi$ and $\chi'$, where the mass difference between the two states is given by $\delta=m_{\chi}'-m_{\chi}>0$. This produces a kinematic suppression for the interaction, where the value of $\delta$ constrains the target masses that will be sensitive to DM scattering. Herein lies the allure of this model. By carefully constraining $m_{\chi}$ and $\delta$, lighter targets such as fluorine can be left blind to these interactions, explaining the lack of a signal at experiments like PICASSO \cite{Behnke2016}. In particular, the case of proton-philic spin dependent inelastic DM (pSIDM) \cite{Kang2019,Kang2019a,DelNobile:2015lxa} also constrains the interaction of DM with targets that have an odd number of neutrons - thus also explaining the absence of a signal at Ge and Xe target experiments such as CDMS and  XENON1T \cite{Ahmed:2009,xenon}. This solution, even though it might require some degree of fine tuning of the operator coefficients, the DM mass and mass splitting, cannot be excluded by the present experimental landscape. Such a model, however, can feasibly be observed other NaI detectors.\\

In this paper we present the best fits to data and explore the sensitivity of SABRE using the three pSIDM models proposed in Ref. \cite{Kang2019} as benchmarks. We will use different velocity distributions to investigate their influence on both fitting models and experimental sensitivity. The paper proceeds by briefly describing in Sec. \ref{sec:sabre} the SABRE experiment, currently in its proof-of-principle stage, then presenting an overview of the rate calculation for interactions assuming different velocity distributions in Sec. \ref{sec:ratecalc}. Our results for fits to the most recent DAMA data and the sensitivity of SABRE to the models investigated are given in Sec. \ref{sec:results}, followed by our conclusions in Sec. \ref{sec:conclusions}.

\section{The SABRE Experiment}
\label{sec:sabre}

\begin{figure}[h!]
    \centering
    \includegraphics[scale=0.35]{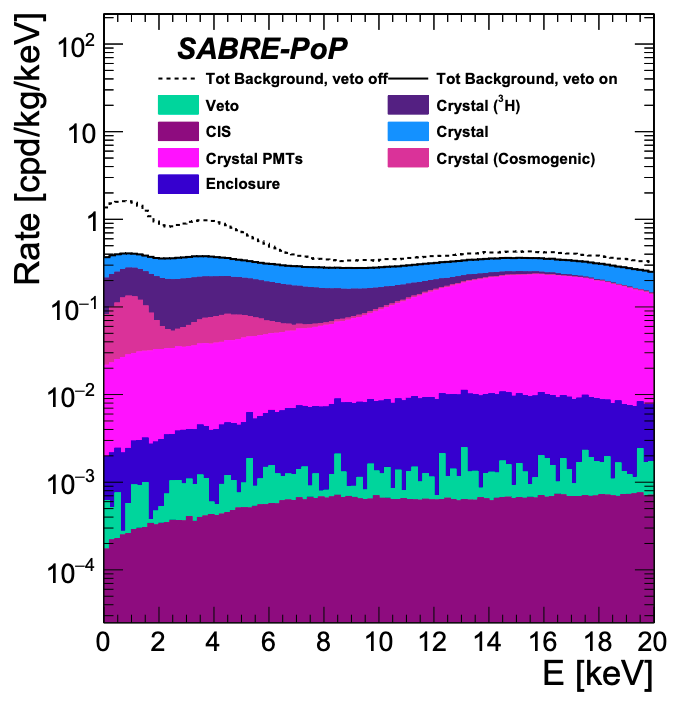}
    \caption{SABRE background from Monte Carlo simulations, taken from \cite{SABREpop}.}
    \label{fig:mc_background}
\end{figure}

The SABRE experiment is a DM DD experiment using a NaI(Tl) target that aims to observe the annual modulation reported by DAMA \cite{SABREpop,Antonello_2019}, and thus confirm or refute the DM claim. It differs in two key ways from DAMA and other NaI DM experiments to ascertain whether the modulation is a genuine DM signal or some yet unaccounted for background. The first is the use of an active veto, producing a much lower background, and the second is detectors placed in both the Northern and the Southern hemispheres. These should be able to distinguish between seasonal modulation (for which detectors in different hemispheres will be out of phase by six months) and DM modulation (for which the detectors will have exactly the same phase). The background computed from Monte Carlo simulations was reported in Ref. \cite{SABREpop}, where it was shown that the application of the active veto greatly reduced the SABRE background, and increased experimental sensitivity to lower recoil energies. For our analysis here, we assume the background with the veto on, shown in Fig. \ref{fig:mc_background} with a solid black line. From the published values, we assume a value of 0.36 cpd/kg/keV within the region of interest.

We will also assume the exposure for the fully instrumented SABRE, which will have 14 NaI crystals of around 3.5 kg each, giving a total active mass of 50 kg. The threshold and detection efficiency are the same as that reported for DAMA/LIBRA in \cite{Bernabei2008}. It should be noted that in reality, these values may differ for the actual SABRE experiment in the future.\\
Ultimately, one detector will be placed in the Stawell Underground Physics Lab in Victoria, Australia and the other in the Gran Sasso National Laboratory in Italy - both of which have a water equivalent depth of 3 km. The present purity levels, shown in Tab. \ref{tab:crystal_purity} suggest that the background levels in Fig. \ref{fig:mc_background} will likely decrease further as technology develops for the full detector, making SABRE the lowest background NaI experiment in the recoil energy range of interest, 1-6 keV.

\begin{table}[!t]
\centering
\begin{tabular}{@{}lcccc@{}}
\toprule
~Experiment & $^{39}$K (ppb) & $^{238}$U (ppt) & $^{232}$Th (ppt) & $^{210}$Pb (mBq/kg) \\ \midrule
~DAMA/LIBRA \cite{Bernabei2008} & $13$ & $0.7-10$ & $0.5-7.5$ & $(30-50)\times 10^{-3}$ \\
~ANAIS-112 \cite{Amare2019} & $32$ & $<0.81$ & $0.36$ & $1.53$ \\
~COSINE-100 \cite{Adhikari} & $35.1$ & $<0.12$ & $<2.4$ & $1.74$ \\
~SABRE (NaI-033) \cite{Suerfu2019} & $4.3$ & $0.4$ & $0.2$ & $0.34$ \\ \bottomrule
\end{tabular}
\caption{Purity levels of NaI crystals of various experiments.}
\label{tab:crystal_purity}
\end{table}
%
%
\section{Dark Matter Recoil Rate}
\label{sec:ratecalc}

\subsection{Dark Matter Interactions}
\label{sec:interactions}

DM interactions with SM particles can be expressed using an Effective Field theory (EFT) with an effective Hamiltonian constructed from a number of operators $\mathcal{O}_j$ that depend on the exact process of the scattering:
\begin{equation}
    \mathcal{H}(\textbf{r})
    =\sum_{\tau=0,1}\sum_{j=1}^{15}c_{j}^{\tau}\mathcal{O}_{j}(\textbf{r})t^{\tau},
\label{hamiltonian}
\end{equation}
where $t^0=\mathbb{I}$ and $t^1=\sigma_3$, the third Pauli matrix. These operators depend on a number of different factors, including the exchanged momentum $\vec{q}$, the incoming relative velocity $\vec{v}$, and the DM and nuclear spins $\vec{j}_{\chi}$ and $\vec{j}_N$. The superscript $\tau$ allows for isoscalar ($\tau=0$) and isovector ($\tau=1$) couplings, which are related to proton and neutron couplings $c^p_j$ and $c^n_j$ via 
\begin{equation}
\begin{split}
    c^p_j &= \frac{1}{2}\left( c^0_j+c^1_j\right),\\
    c^n_j &= \frac{1}{2}\left(c^0_j-c^1_j\right).\\
\end{split}
\end{equation}
 Following the methodology of Ref. \cite{Liam2013,Anand:2013yka,Cirelli2013}, these operators can be used to calculate the cross section for scattering between DM and a nucleus via typical EFT formalism.\\
The couplings $c_i$ influence the detection rate via the inclusion of the nuclear form factors $F^{(ab)}_{ij}(v,q)$, which convert the scattering cross section of a single nucleon into a cross section that can be used for a full target nucleus. This is computed using the squared scattering matrix element, averaged over spins;
\begin{equation}
    \frac{1}{2j_{\chi}+1}\frac{1}{2j_{T}+1} \sum_{\text{spins}}|\mathcal{M}|^2=
    \sum_{i,j}^{15}\sum_{a,b=0,1} c_i^{(a)}c_j^{(b)}F^{(ab)}_{ij}(v,q),
\label{og_cross}    
\end{equation}
where $c_{i,j}$ are the same coefficients of Eq. \ref{hamiltonian}. Here, and throughout the paper, $N$ subscripts refer to an individual nucleon while a $T$ refers to the full target nucleus. A full list of the form factors used in this analysis for $^{23}$Na and $^{127}$I can be found in Appendix A of Ref. \cite{Liam2013}.\\
Ref. \cite{Kang2019} presented analysis for DM scattering off a NaI nucleus with $j_{\chi}=0,~1/2,$ and$~1$ in an attempt to find a combination of these coupling constants that is able to fit both the annual modulation observed by DAMA, and the absence of a signal at other experiments. To ease computation, in the process they expressed these coupling constants as a vector $\boldsymbol{c_0}$
\begin{equation}
    \boldsymbol{c_0} = \sum_{i}^{15}\sum_{a=0,1} \boldsymbol{c}_i^{(a)} \hat{e}_i^{a} = c_0 \boldsymbol{\hat{c}_0},
    \label{coupling_vector}
\end{equation}
where $e_i^{(a)}$ are unit vectors. In this way, the fit of the couplings constants can be separated into the fit of the direction and norm of the vector $\boldsymbol{\hat{c}_0}$. Thus, all the information about the relative contributions of different operators/form factors are contained within $\boldsymbol{\hat{c}_0}$, allowing for $c_0$ to be pulled out of Eq. \ref{og_cross} as a common factor:
\begin{equation}
    \frac{1}{2j_{\chi}+1}\frac{1}{2j_{T}+1} \sum_{\text{spins}}|\mathcal{M}|^2=
    c_0^2\sum_{i,j}\sum_{a,b=0,1} \hat{c}_i^{(a)}\hat{c}_j^{(b)}F^{(ab)}_{ij}(v,q).
\label{full_cross}    
\end{equation}
Adopting the parameterisation 
\begin{equation}
    \sigma_0 \equiv \frac{\mu^2_N c^2_0}{\pi}
\end{equation}
then allows for fits to a DM cross section $\sigma_0$, as well as $m_{\chi}$, $\delta$, and the components of $\boldsymbol{\hat{c}_0}$. Fits from Ref. \cite{Kang2019} for the lowest tension pSIDM models in the three different spin cases, assuming an escape velocity of $550$ ms$^{-1}$, are shown in Tab. \ref{couplings}.\\

\begin{table}[!h]
\centering
\begin{tabular}{@{}ccccccc@{}}
\toprule
~Case & Spin ($j_{\chi}$) & $m$ (GeV) & $\sigma_0$ (cm$^2$) & $\delta$ (keV) & \multicolumn{2}{c}{Non zero $\boldsymbol{\hat{c}_0}$ components} \\ \midrule
1 & 0 & $11.1$ & $3.9\times10^{-27}$ & $22.8$ & $\hat{c}^0_7 = 0.68$ & $\hat{c}^1_{7} = 0.73$ \\ \midrule
2 & 1/2 & $11.6$ & $4.7\times10^{-28}$ & $23.7$ & \begin{tabular}[c]{@{}c@{}}$\hat{c}^0_4 = -0.0014$\\ $\hat{c}^0_5 = -0.032$\\ $\hat{c}^0_6 = 0.692$\end{tabular} & \begin{tabular}[c]{@{}c@{}}$\hat{c}^1_4 = -0.0015$~\\ $\hat{c}^1_5 = -0.0166$~\\ $\hat{c}^1_6 = 0.7217$~\end{tabular} \\ \midrule
3 & 1 & $11.4$ & $5.7\times10^{-32}$ & $23.4$ & \begin{tabular}[c]{@{}c@{}}$\hat{c}^0_4 = 0.0717$\\ $\hat{c}^0_5 = 0.1892$\end{tabular} & \begin{tabular}[c]{@{}c@{}}$\hat{c}^1_4 = 0.0753$~\\ $\hat{c}^1_5 = 0.9764$~\end{tabular} \\ \bottomrule
\end{tabular}
\caption{The non zero coupling constants and fit to DAMA/LIBRA data from Ref. \cite{Kang2019} analysis.}
\label{couplings}
\end{table}

To demonstrate the proton-philic nature of these three models, we also present the explicit proton and neutron couplings in Tab. \ref{tab:proton_neutron_coupling}. Neutron couplings are around an order of magnitude weaker than the proton ones, demonstrating why these models are able to evade detection by the XENON collaboration.\\
Using these coupling constants, we are able to assess SABRE's sensitivity to the three models for a range of $m_{\chi}$ and $\sigma_0$, as well as finding the best fit to DAMA/LIBRA for $m_{\chi}$, $\sigma_0$, and $\delta$ (keeping the direction of the coupling `vector' fixed) and comparing it to the lowest tension fit computed by Kang et. al. \cite{Kang2019}.

\begin{table}[!t]
\centering
\begin{tabular}{@{}ccc@{}}
\toprule
~Case               & Proton coupling        & Neutron coupling~               \\ \midrule
1                  & $\hat{c}^p_7=1.41$    & $\hat{c}^n_7=-0.05$           \\ \midrule
\multirow{3}{*}{2} & $\hat{c}^p_4=-0.0029$ & $\hat{c}^n_4=1.0\times10^{-3}$ \\
                   & $\hat{c}^p_5=-0.0486$  & $\hat{c}^n_5=-0.0154$          \\
                   & $\hat{c}^p_6=1.414$    & $\hat{c}^n_6=-0.0298$          \\ \midrule
\multirow{2}{*}{3} & $\hat{c}^p_4=0.147$   & $\hat{c}^n_4=-0.0036$          \\
                   & $\hat{c}^p_5=1.166$    & $\hat{c}^n_5=-0.788$           \\ \bottomrule
\end{tabular}
\caption{The proton and neutron couplings for the models shown in Table \ref{couplings}.}
\label{tab:proton_neutron_coupling}
\end{table}
\subsection{Benchmark Models}
\label{sec:benchmodels}

The exact interaction that occurs between DM and a nucleus is described by the non-relativistic nucleon operators $\mathcal{O}_i$ corresponding to the non-zero couplings $c_i$ for a particular model. Each of these operators will correspond to an effective high energy operator - the result of integrating out the (unknown) mediator, which tells us about the DM model under consideration. In some cases, a non-relativistic nucleon operator may be associated with more than one high energy effective operator, meaning that the two models cannot be distinguished via direct detection experiment. All operators given in this section to describe the benchmark models from Ref. \cite{Kang2019} under consideration are non-relativistic and, apart from $\mathcal{O}_4$, all are suppressed by powers of $v$ and/or $q$, each of which approximately gives a suppression on the order of $(10^{-3})$. For more detail as to how these are derived from their corresponding high energy effective operators, see Ref. \cite{Liam2013,Cirelli2013}.\\  
Case 1 depends on operator $\mathcal{O}_7$, and as such describes a nucleon spin ($s_n$) dependent interaction with explicit velocity dependence:
\begin{equation}
    \mathcal{O}_7 = \boldsymbol{s}_N\cdot\boldsymbol{v}^{\perp}
\end{equation}
which will have a suppression factor of $v^2$. Here, as in Ref. \cite{Kang2019}, $\boldsymbol{v}^{\perp}$ is defined as  $v^{\perp}=\vec{v}+\frac{\vec{q}}{2\mu_{N\chi}}$, satisfying $v^{\perp}\cdot \vec{q} = 0$ and $\left(v^{\perp}\right)^2=v^2-v_{min}^2$ where in the inelastic case
\begin{equation}
    v_{min}=\frac{1}{\sqrt{2m_TE_R}}\left|\frac{m_TE_R}{\mu_{\chi,T}}+\delta\right|.
\label{min_vel}    
\end{equation}

Combinations of operators 4, 5, and 6 dictate cases 2 and 3, so these are expected to produce similar DM interactions. All three depend on the DM spin ($s_{\chi}$), and $\mathcal{O}_5$ and $\mathcal{O}_6$ will have either $q^4$ or $q^2v^2$ momentum/velocity suppression, while $\mathcal{O}_4$ is the standard spin-dependent operator, not suppressed by any power of $q$ or $v$. Operator $\mathcal{O}_5$ does not depend on the nucleon spin $\boldsymbol{s}_N$, and therefore is spin-independent, but still suppressed by $v^2q^2$,
\begin{equation}
\begin{split}
    \mathcal{O}_4 &= \boldsymbol{s}_{\chi}\times\boldsymbol{s}_{N},\\
    \mathcal{O}_5 &= i\boldsymbol{s}_{\chi}\cdot\left(\boldsymbol{q}\times \boldsymbol{v}^{\perp}\right),\\
    \mathcal{O}_6 &= \left(\boldsymbol{s}_{\chi}\cdot\boldsymbol{q}\right)\left(\boldsymbol{s}_{N}\cdot\boldsymbol{q}\right).\\
\end{split}    
\end{equation}
Models two and three will include additional interference terms due to the fact that $F_{4,5},F_{4,6}\neq 0$.
The momentum suppression present in all three models considered here is to be expected, as this alleviates the constraints implied by droplet detectors and bubble chambers \cite{Scopel_2015}. As such any model that reduces tension between DAMA/LIBRA and other experiments is likely to include momentum suppression to some degree. 

%
%
\subsection{Differential Average and Modulated Rates}
\label{sec:diffrates}

The differential interaction rate, with respect to nuclear recoil energy $E_R$, between a target nucleus and DM particle is given by
\begin{equation}
    \frac{dR}{dE_R} = N_T\frac{\rho}{m_{\chi}} \int v f_{lab}(\vec{v})\frac{d\sigma_T}{dE_R}d^3v,
\label{eq:general_rate}    
\end{equation}
where $N_T$ is the number of target atoms per kg of target, $f_{lab}(v)$ is the DM velocity distribution in the lab frame, and the velocity integral goes from $v_{min}$ in the lab frame, up to the galaxy escape velocity. We can express the differential cross section as
\begin{equation}
\begin{split}
    \frac{d\sigma_T}{dE_R} 
        &= \frac{m_T}{2\pi v^2} 
            \left[\frac{1}{2j_{\chi}+1}\frac{1}{2j_{T}+1}     
            \sum_{\text{spins}}|\mathcal{M}|^2\right],\\
        &= \frac{m_T}{2v^2}\frac{\sigma_0}{\mu^2_{N}}
            \left[\sum_{i,j}\sum_{a,b=0,1} \hat{c}_i^{(a)}\hat{c}_j^{(b)}F^{(ab)}_{ij}(v,q)\right].\\    
\end{split}    
\label{idm_rate_full}  
\end{equation}
Thus, to evaluate an experiments sensitivity, or to fit to $m_{\chi}$, $\sigma_0$, and $\delta$, a particular velocity distribution $f(v)$ and the direction of the vector $\boldsymbol{\hat{c}_0}$ must be chosen in order to compute the interaction rate. Typically, one can interpret Eq. \ref{idm_rate_full} as the particle physics content of the interaction rate, while $f(v)$ is the astrophysical contribution.
To allow for easier computation by separating the two, we make the observation that all the terms in the form factor sum are either independent of velocity, or proportional to $v^2$, given in appendix A.2. This allows us to write
\begin{equation}
    F^{(ab)}_{ij}(v,q) = F^{(ab),1}_{ij}(q)+v^2 F^{(ab),2}_{ij}(q),
\end{equation}
and therefore separate the cross section into 2 terms, with different velocity dependance.
\begin{eqnarray}
\frac{d\sigma_T}{dE_R} 
        &=& \frac{1}{v^2}\left(\frac{d\sigma_T^1}{dE_R} + v^2 \frac{d\sigma_T^2}{dE_R}\right)\\
\frac{d\sigma_T^l}{dE_R} 
        &=& \frac{m_T}{2}\frac{\sigma_0}{\mu^2_{N}}
            \left[\sum_{i,j}\sum_{a,b=0,1} \hat{c}_i^{(a)}\hat{c}_j^{(b)}F^{(ab),l}_{ij}(q)\right].
\end{eqnarray}

Using this, we are able to rewrite Eq. \ref{eq:general_rate} in terms of two integrals: 
\begin{equation}
\frac{dR}{dE_R} = N_T\frac{\rho}{m_{\chi}} \frac{\sigma_0m_T}{2\mu^2_{N}} \sum_{i,j}\sum_{a,b=0,1} \hat{c}_i^{(a)}\hat{c}_j^{(b)} \left(F^{(ab),1}_{ij}(q)\int \frac{f_{lab}(\vec{v})}{v}d^3v+F^{(ab),2}_{ij}(q)\int v f_{lab}(\vec{v})d^3v\right).
\end{equation}
These can be computed after expressing the DM velocity distribution in the lab in terms of the DM velocity distribution in the galaxy frame $f(v)$
\begin{eqnarray}
f_{lab}(\vec{v}) &=& f(|\vec{v}-\vec{v}_E|),
\end{eqnarray}
where $\vec{v}_E=\vec{v}_\odot+\vec{v}_t$ is the Earth's velocity taking into account the solar velocity $\vec{v}_{\odot}$ and the rotation of the Earth around the Sun $\vec{v}_t$. Thus the velocity integrals can be expressed as

\begin{equation}
\begin{split}
   \int \frac{f_{lab}(\vec{v})}{v}d^3v= g(v_{min})&=\iint_{\mathcal{D}}v~f_{lab}(\vec{v})dv~d\Omega,\\   
    \int v f_{lab}(\vec{v})d^3v= h(v_{min})&=\iint_{\mathcal{D}}v^3~f_{lab}(\vec{v})dv~d\Omega.\\
\end{split}
\label{vel_ints}
\end{equation}
with $\mathcal{D}$ defined as
\begin{equation}
    v>v_{min}(E_R),\quad |\vec{v}-\vec{v}_E|<v_{esc}.
\end{equation}

These velocity integrals then form prefactors that, aside from $v_{min}$, do not depend on the particle physics DM model in question. They are then multiplied by the appropriate form factors, giving
\begin{eqnarray}
    \frac{dR}{dE_R} &=& N_T\frac{\rho}{m_{\chi}}\left[\frac{d\sigma_1}{dE_R}g(v_{min})+\frac{d\sigma_2}{dE_R}h(v_{min})\right]\\
    &=& N_T \frac{\rho}{m_{\chi}} \frac{\sigma_0 m_T}{2\mu^2_{N}} \sum_{i,j}\sum_{a,b=0,1} \hat{c}_i^{(a)}\hat{c}_j^{(b)} \left(F^{(ab),1}_{ij}(q)g(v_{min})+F^{(ab),2}_{ij}(q)h(v_{min})\right).
\end{eqnarray}
where $q$ is related to $E_R$ by
\begin{equation}
    q^2 = 2m_TE_R.
\end{equation}
 The benefit of expressing the rate in this way is that it allows us to separately calculate the astro and particle physics contributions. This makes computation and comparison for different combinations of DM interaction models and velocity distributions significantly easier to perform, as it removes the need to reevaluate these integrals for every different DM model.

\subsubsection*{Modulating Signal} 
Due to the rotation of the Earth around the Sun, its velocity relative to the galactic DM will take the form of a cosine function with a period of one year. As such, the total interaction rate is expected to follow the same distribution. To make this clear, and to separate the average and modulating components, expressions in Eq. \ref{vel_ints} can be projected onto $A+B\cos[\omega(t-t_0)]$, giving an interaction rate of the form
\begin{equation}
\begin{split}
    \frac{dR(t)}{dE_R}&=\frac{dR_0}{dE_R}+\frac{dR_m}{dE_R}\cos[\omega(t-t_0)]\\
    \Rightarrow R(t)&=R_0\left(1+\alpha\cos[\omega(t-t_0)]\right).
\end{split}
\end{equation}
Here $\alpha=R_m/R_0$ is the modulation amplitude, and for most DM velocity distributions is expected to be on the order of 1\% \cite{Patrignani:2016xqp}. Observation of this modulating signal is thought to be a clear signpost of DM within the galaxy, and can be observed without needing to assume any particular DM interaction model.
In addition to this, pSIDM models in particular are expected to have a much stronger modulation than standard elastic WIMP models \cite{Tomar:2019urz}. Thus, analysis of a clear $R_0$ attributable to DM as well as $R_m$ may help to distinguish between various models under consideration.

\subsection{Dark Matter Velocity Distributions}
\label{sec:veldistrib}
The velocity distribution typically assumed for galactic DM is the Standard Halo Model (SHM), where the DM follows a Maxwell Boltzmann distribution
\begin{equation}
    f_{SHM}(v)=\frac{1}{(\pi v_0^2)^{3/2}}\exp\left[-\frac{1}{v_0^2}\left(\boldsymbol{v}-\boldsymbol{v}_E\right)^2\right].
\end{equation}
The values used for these constants in this analysis are given in Appendix A.1.\\
Recently, however, results from the \textit{Gaia} satellite and astrophysical simulations have suggested that the SHM is too simplistic to describe the DM content of the Milky Way \cite{Patrignani:2016xqp,stream_dist}. There are a large number of new halo models that are now being considered, some of which may change the interpretation of data gleaned from direct detection experiments. One such model accounts for the substructure from the tidal stream disruption of satellite galaxies of the Milky Way, a stream S1 associated with DM that ``hits the Solar system slap in the face'' \cite{stream_dist}. This anisotropic substructure can be accounted for by adding terms to the SHM distribution, forming a distribution we will refer to SHM+Str. These additional terms take the form
\begin{equation}
    f_{Str}(v)=\frac{1}{(8\pi^3\boldsymbol{\sigma}^2)^{1/2}}\exp\left[-\left(\boldsymbol{v}-\boldsymbol{v}_E+\boldsymbol{v}_{Str}\right)^T\frac{\boldsymbol{\sigma}^{-2}}{2}\left(\boldsymbol{v}-\boldsymbol{v}_E+\boldsymbol{v}_{Str}\right)\right],
\label{stream}
\end{equation}
where the dispersion tensor $\boldsymbol{\sigma}$ is diagonal when derived in cylindrical coordinates, given by $\boldsymbol{\sigma}^2=\text{diag}(\sigma_r^2,\sigma_{\phi}^2,\sigma_z^2)$.

These terms are then combined with the SHM as a fraction of the local density, so
\begin{equation}
\begin{split}
    f_{SHM+Str}(v) &= \left(1-\frac{\rho_s}{\rho}\right)f_{SHM}(v)+\frac{\rho_s}{\rho}f_{Str}(v),\\
\end{split}    
\end{equation}
where $\rho$ and $\rho_s$ are the relative population density of the SHM and stream distributions, usually defined so that around 10\% of the DM is in the stream.
Again, the particular values used for this distribution in our analysis are included in Appendix A.1.

\subsection{Detector Response}
\label{sec:detectorresp}
The expression given in Eq. \ref{idm_rate_full} makes the implicit assumption that we are dealing with an idealised detector with 100\% detection efficiency. In reality, the detection process will introduce additional threshold cutoffs, and require calibration between the actual nuclear recoil energy, and the energy measured by the detector.\\

\subsubsection*{Quenching factors}
Both experiments of interest for this analysis, DAMA and SABRE, are scintillation detectors, and thus the quenching factor $Q$ needs to be accounted for. This is used to equate the light output of an electron (what is actually detected by equipment) with the nuclear recoil of the Na or I nucleus (the result of the DM scattering). Essentially, it is a unit conversion between the observed electron equivalent energy $E_{ee}$ (keV$_{ee}$) and the actual nuclear recoil energy $E_R$ (keV$_{nr}$). This correction takes the form
\begin{equation}
\begin{split}
        E_R&=\frac{E_{ee}}{Q},\\
    \frac{dR}{dE_{ee}}&=\frac{dR}{dE_R}\frac{dE_R}{dE_{ee}}.
\end{split}
\end{equation}
The commonly accepted quenching factor of I is a constant 0.09, and for this analysis, the value used for Na's quenching factor is the constant $Q=0.3$ assumed by Ref. \cite{Bernabei2018,Kang2019}. Analysis conducted in Ref. \cite{shields} and ongoing at the Australian National University has demonstrated that in reality Na's quenching factor is energy dependent, effectively shifting the peak in the modulating interaction to lower energies. 
Initial tests were conducted to find a fit to the DAMA/LIBRA data using these new measurements for the Na quenching factor, but due to the shift in peak location, upon assuming the couplings in Tab. \ref{couplings} the interaction rate did not fit the trend suggested by the data. It is likely that in order to use the quenching factor given in Ref. \cite{shields}, new analysis of the kind presented in Ref. \cite{Kang2019} will be required to find more appropriate coupling constants. It is possible that this consideration may help to alleviate tension between various results.

\subsubsection*{Efficiency and resolution}
The threshold detection efficiency will influence the probability of an event of a given energy actually being observed by the detector. For this analysis of SABRE, the values reported by DAMA/LIBRA are sufficient. This efficiency rises linearly from a value of 0.70 at 1 keV$_{ee}$ up to a value of 1 at 8 keV$_{ee}$ and above \cite{Bernabei_highQE}.

In addition to this, the energy resolution of each detector will influence the observed rate of interaction, effectively smearing the signal and causing recoils of energy $E_{ee}$ to be observed as a Gaussian distributed spectrum \cite{Lewin1996}. Thus, the differential rate will undergo a transformation
\begin{equation}
    \frac{dR}{dE'}=\frac{1}{(2\pi)^{1/2}}\int_{0}^{\infty}\frac{1}{\Delta E_{ee}}\frac{dR}{dE_{ee}}\exp\left[\frac{-(E'-E_{ee})^2}{2(\Delta E_{ee})^2}\right]dE_{ee},
\label{ob_diff}    
\end{equation}
where $\Delta E$ is the energy resolution of the detector. This analysis will again use the same expression as DAMA/LIBRA for SABRE, given in Ref. \cite{Bernabei2008} as:
\begin{equation}
   \Delta E = \left(0.0091\frac{E_{ee}}{\keV_{ee}}+0.488\sqrt{\frac{E_{ee}}{\keV_{ee}}}\right)\keV_{ee}. 
\end{equation}
The results are binned with a width of 0.5 keV$_{ee}$ centred around the integers and half integers, going from 1 keV$_{ee}$ up to 5 keV$_{ee}$.

\subsubsection*{Multi element targets}
For DM targets that are made up of more than one element, such as NaI, the calculations for each element must be done separately, then added together. Thus the total, overall rate will be given by the rate of each target nucleus $i$, weighted by their contributing masses $m_i$ as a fraction of the total molecular mass $m_\text{Tot}$:
\begin{equation}
    \frac{dR_\text{Tot}}{dE_R}=\sum_{i}\frac{m_i}{m_\text{Tot}}\frac{dR_i}{dE_R}.    
\end{equation}
So in the case of NaI the observed interaction rate is
\begin{equation}
    \frac{dR_\text{Tot}}{dE_R}= \frac{m_{Na}}{m_{Na}+m_I}\frac{dR_{Na}}{dE_R}+\frac{m_I}{m_{Na}+m_I}\frac{dR_I}{dE_R}.
\end{equation}\\

%
%
\section{Results}
\label{sec:results}

\subsection{Best Fits for the Differential Rates}
\label{sec:bestfits}

\begin{figure}[!t]
    \centering
    \subfloat[Model 1. \label{1_lt}]{%
        \includegraphics[width=0.5\textwidth]{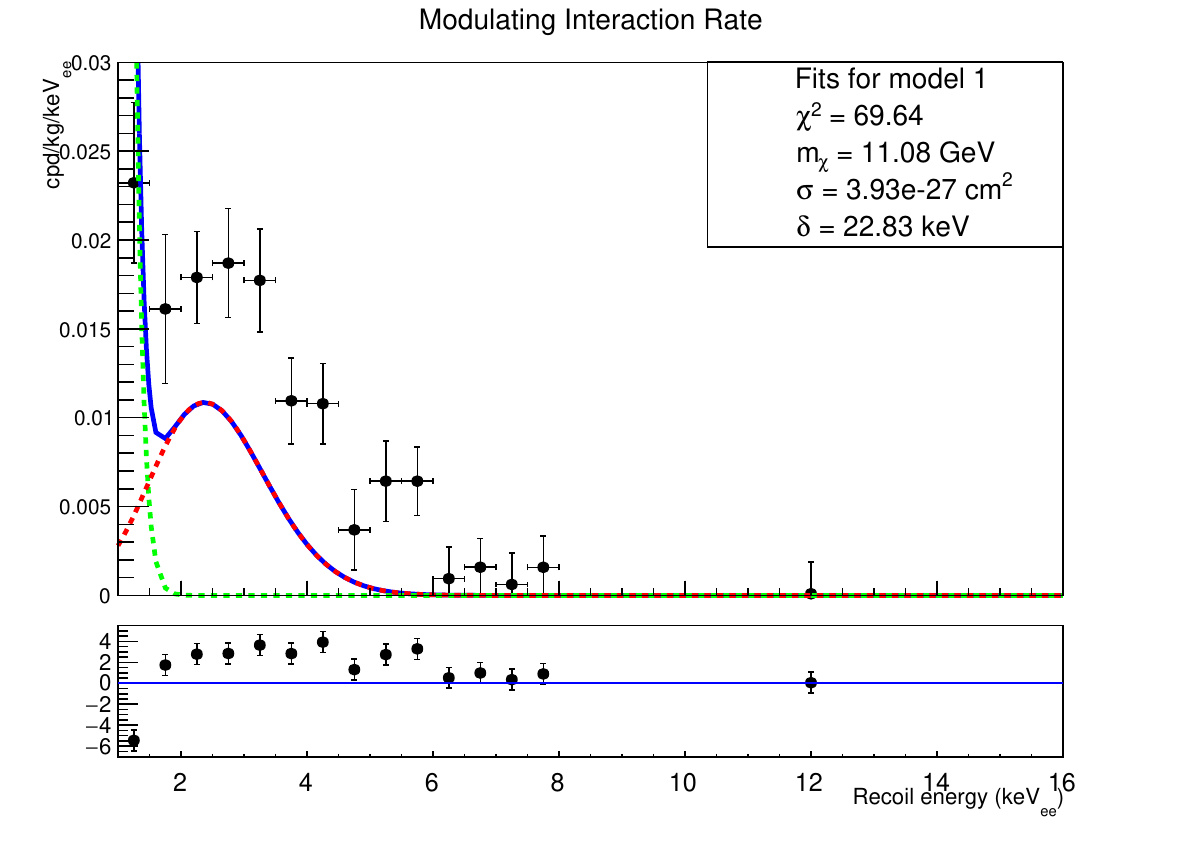}}
    \subfloat[Model 2.\label{2_lt}]{%
        \includegraphics[width=0.5\textwidth]{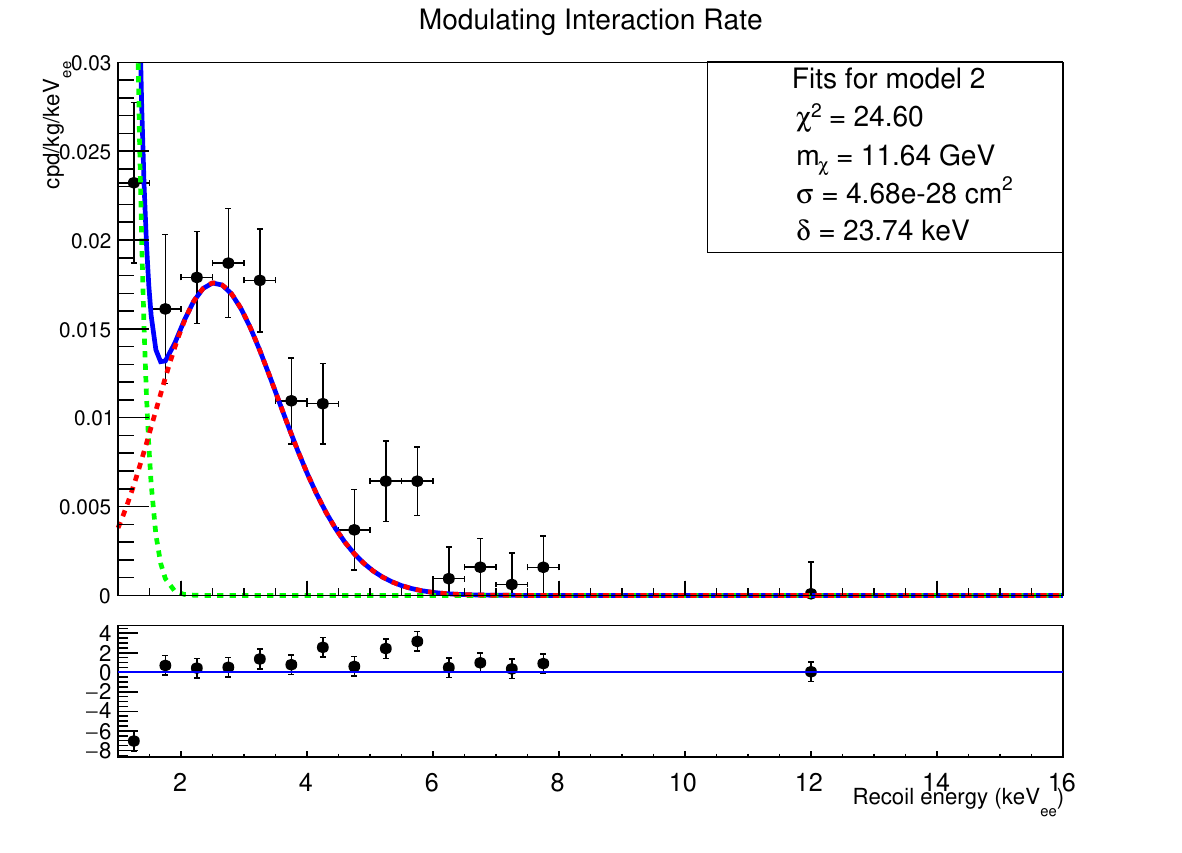}}\\
    \subfloat[Model 3. \label{3_lt}]{%
        \includegraphics[width=0.5\textwidth]{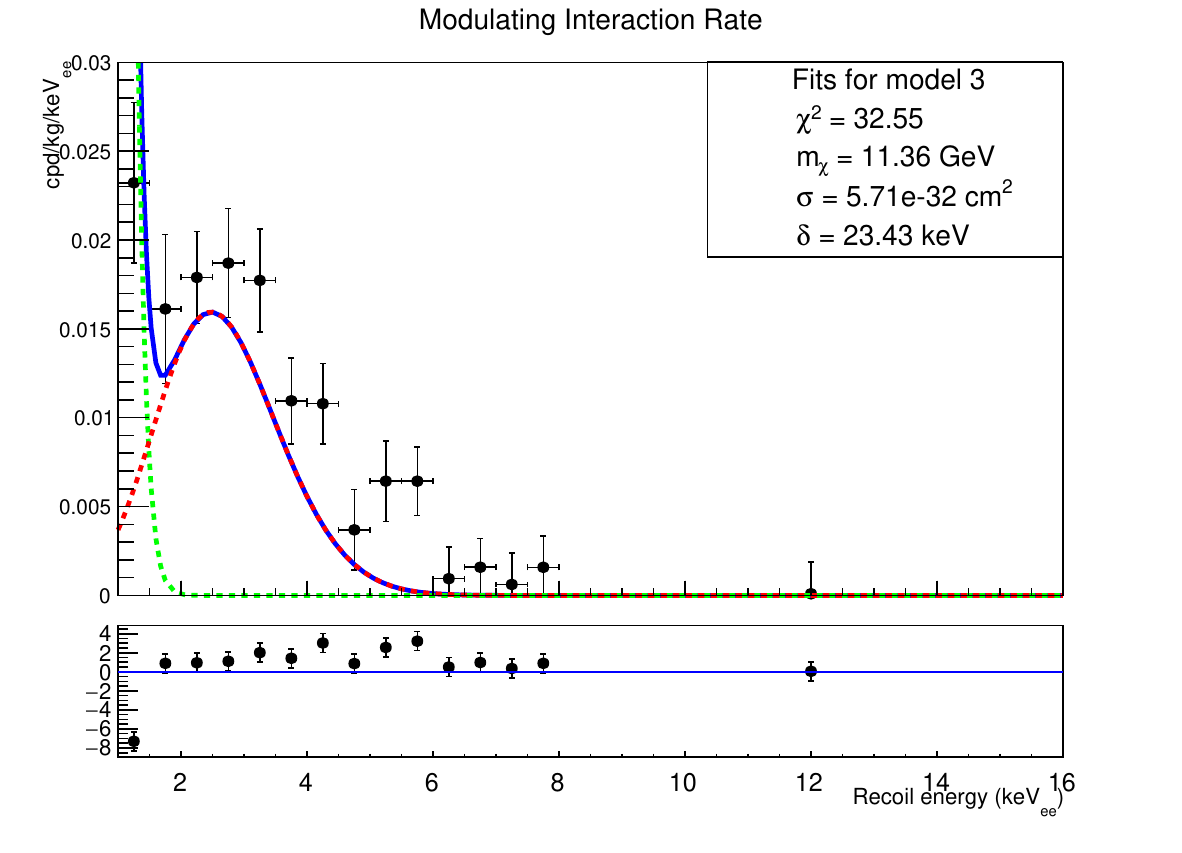}}
    \caption{Lowest tension fits presented in Ref. \cite{Kang2019}. Here the green dotted line gives the iodine interaction rate, the red sodium, and the blue the total overall observed rate.}
    \label{lt_fits}
\end{figure}

\begin{figure}[!ht]
    \centering
    \subfloat{%
        \includegraphics[width=0.5\textwidth]{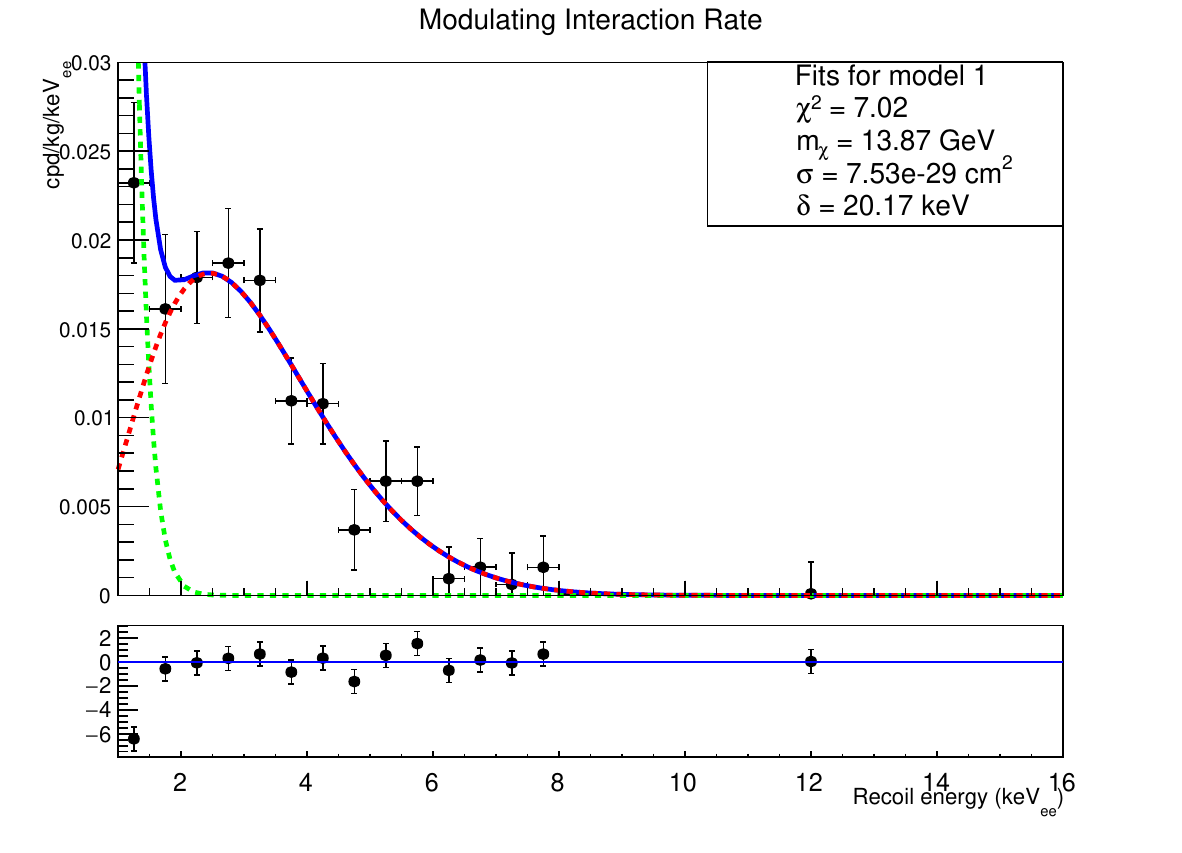}}
    \subfloat{%
        \includegraphics[width=0.5\textwidth]{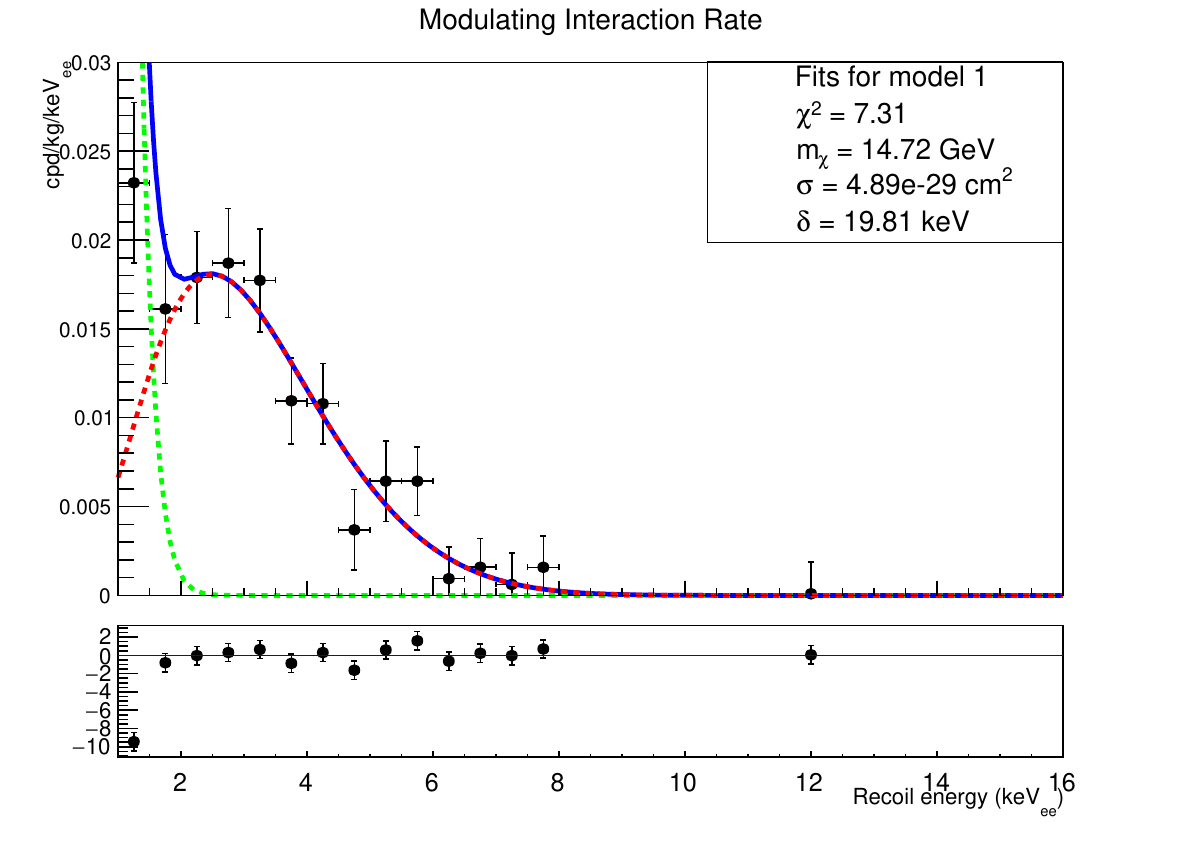}}\\
        \subfloat{%
        \includegraphics[width=0.5\textwidth]{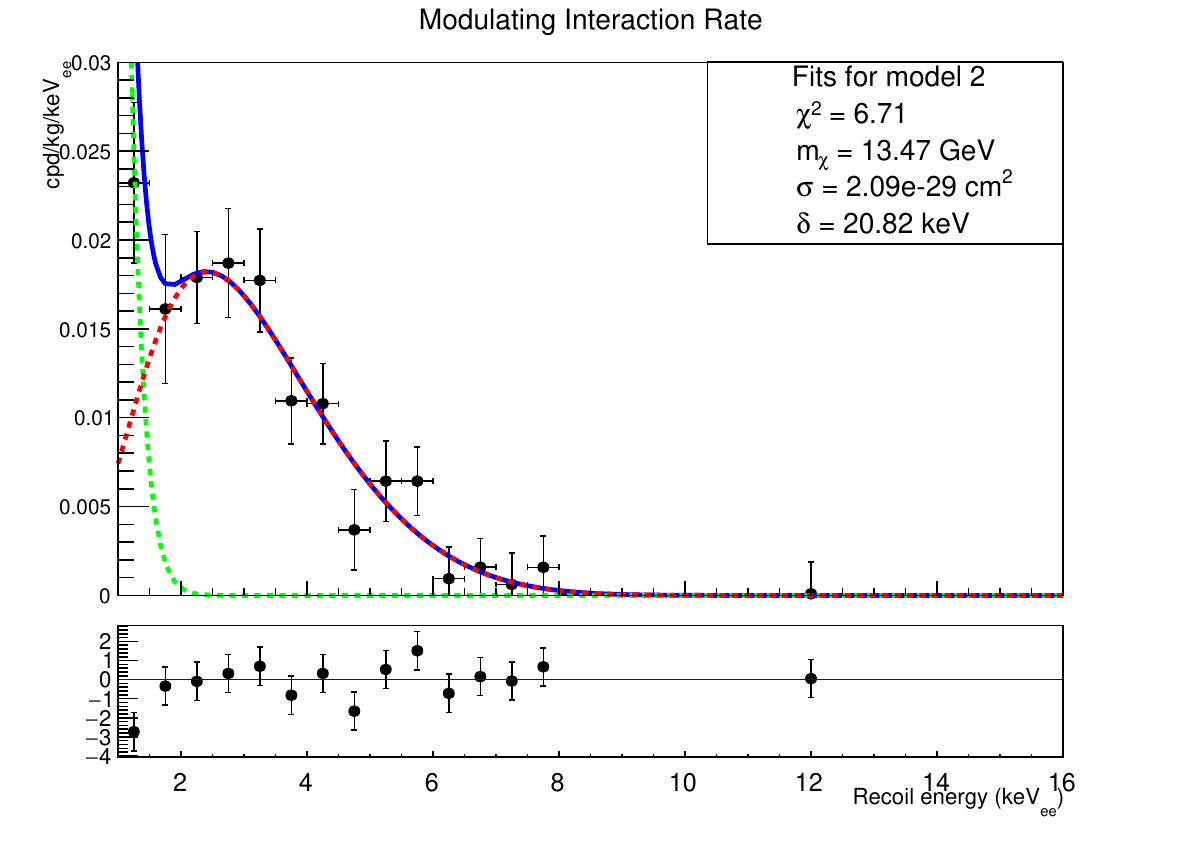}}
    \subfloat{%
        \includegraphics[width=0.5\textwidth]{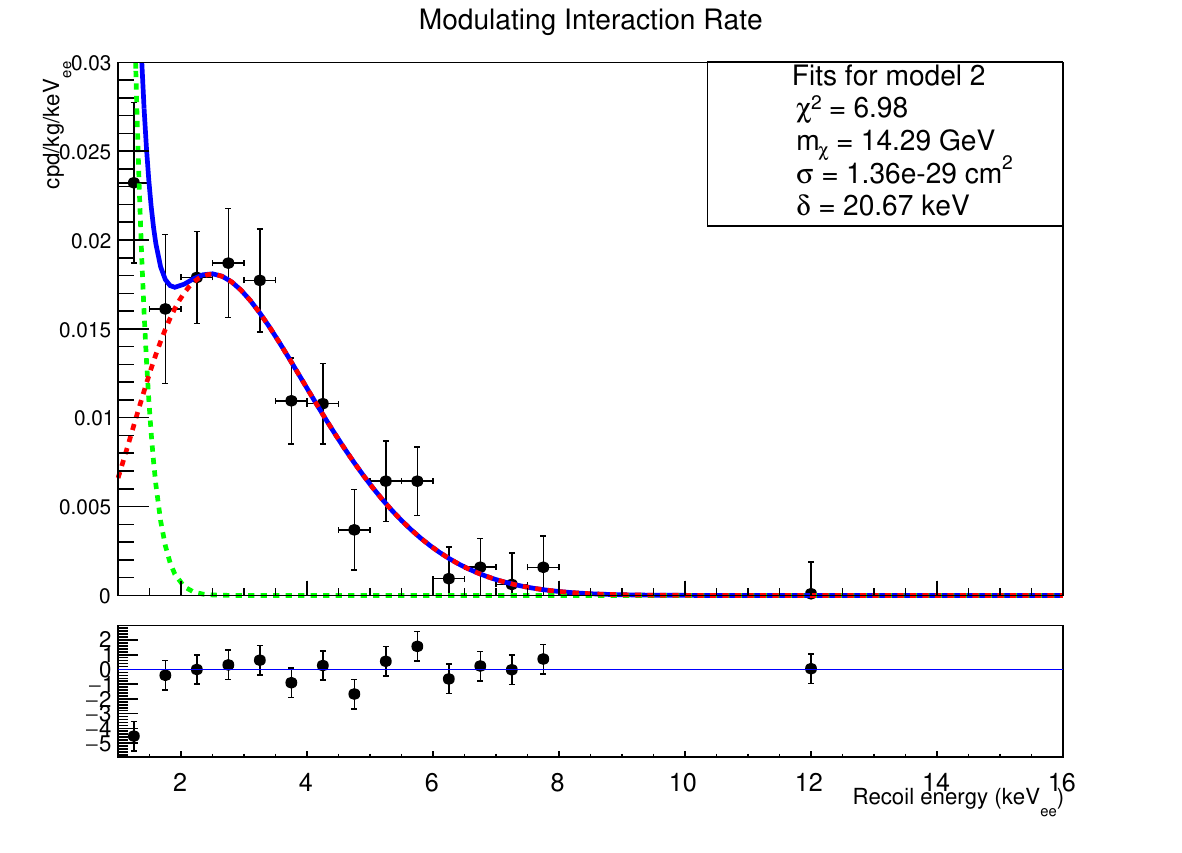}}\\
        \subfloat{%
        \includegraphics[width=0.5\textwidth]{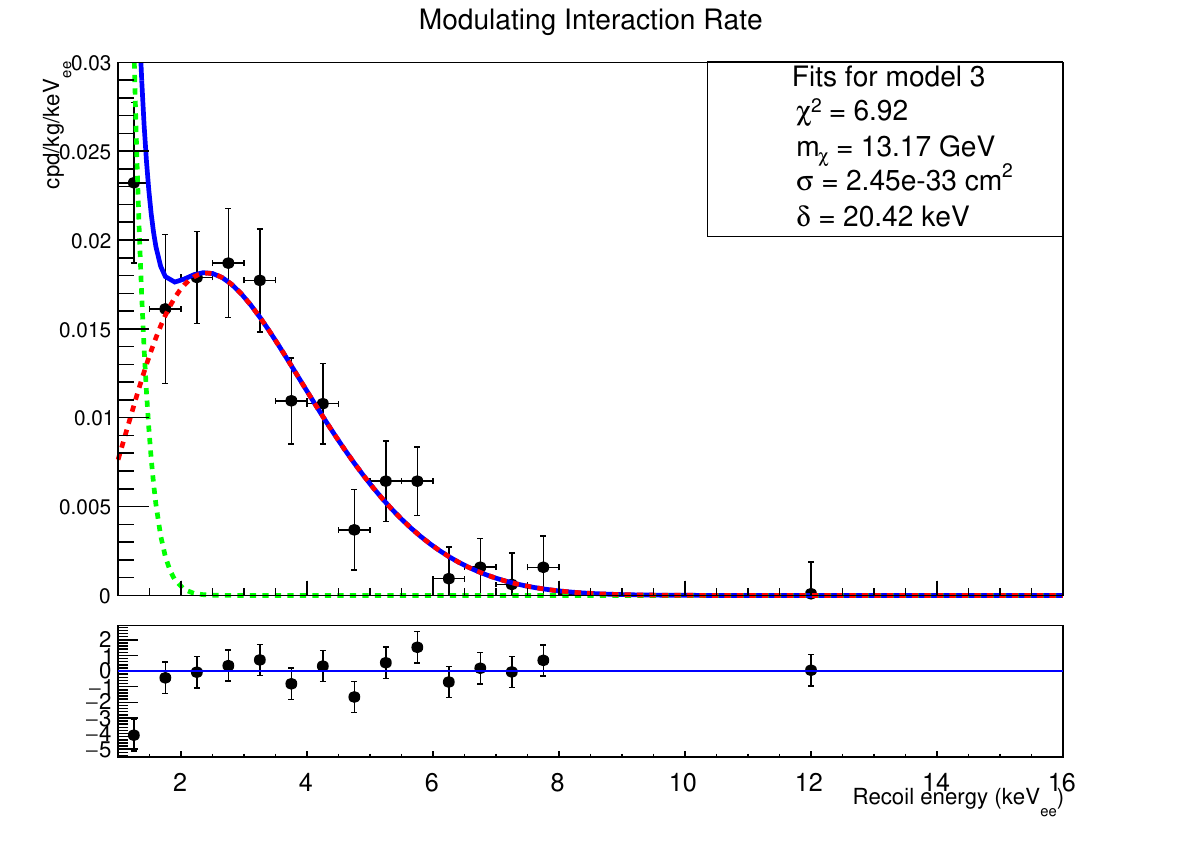}}
    \subfloat{%
        \includegraphics[width=0.5\textwidth]{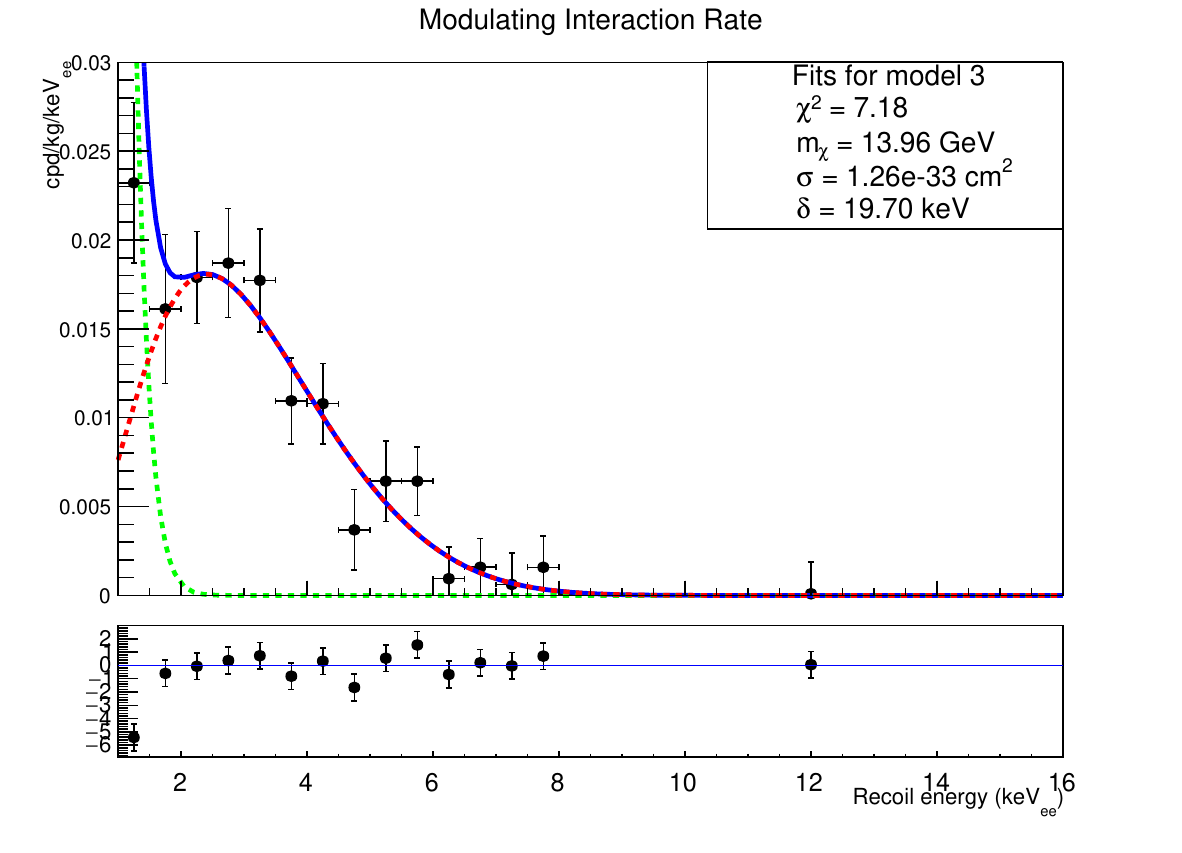}} 
    \caption{Best fits to DAMA/LIBRA data with SHM (left) and SHM+Stream (right), for Model 1 (top), model 2 (center), and model 3 (bottom).}
    \label{best_fit_1}
\end{figure}

As was noted in Ref. \cite{Kang2019} and in Sec. \ref{sec:ratecalc} of this paper, the fits reported in Tab. \ref{couplings} are based on achieving the lowest tension between various experiments, and as such are not necessarily the closest fit to the DAMA/LIBRA data. These low tension fits are reproduced in Fig. \ref{lt_fits} from Ref.\cite{Kang2019}, assuming a constant Na quenching factor of 0.3. Note that the $\chi^2$ values found in this reproduction are slightly different to those previously published, in part due to the slightly different data points (ours taken from Ref. \cite{Bernabei2018}). Although the tension is low for these models, they are not the best fits available to the DAMA/LIBRA data, compared to other models that have been suggested, but then rejected by null results from other experiments \cite{Patrignani:2016xqp}. \\
Using RooFit, we have conducted new fits to the DAMA/LIBRA results for $m_{\chi}$, $\sigma_0$, and $\delta$, assuming the DM spin and couplings given in Tab. \ref{couplings}. Effectively, we are changing only the normalisation of $\boldsymbol{c}_0$ as defined in Eq. \ref{coupling_vector} and not the direction. These were done for both the SHM and SHM+Stream distributions with $Q_{Na}=0.3$. Results are shown in Tab. \ref{tab:fit_results}, and in Fig. \ref{best_fit_1}.

\begin{table}[!h]
\centering
\begin{tabular}{@{}ccccccc@{}}
\toprule
~Velocity distribution~     & ~Model~ & $m_{\chi}$ (GeV) & $\sigma_0$ (cm$^2$)    & $\delta$ (keV) & ~$\chi^2$/dof~ \\ \midrule
\multirow{3}{*}{SHM}        & $1$   & $13.87$          & $7.53\times10^{-29}$ & $20.17$        & $7.02/12$    \\
                            & $2$   & $13.47$          & $2.09\times10^{-29}$ & $20.82$        & $6.71/12$    \\
                             & $3$   & $13.17$          & $2.45\times10^{-33}$ & $20.42$        & $6.92/12$    \\ \midrule
\multirow{3}{*}{SHM+Stream} & $1$   & $14.72$          & $4.89\times10^{-29}$ & $19.81$        & $7.31/12$    \\
                            & $2$   & $14.29$          & $1.36\times10^{-29}$ & $20.67$        & $6.89/12$   \\
                            & $3$   & $13.96$          & $1.26\times10^{-33}$ & $19.70$        & $7.18/12$   \\ 
\bottomrule
\end{tabular}
\caption{Fits to various DM models from the DAMA/LIBRA data.}
\label{tab:fit_results}
\end{table}

In general, these fits have increased the mass and decreased the cross section compared to those in Ref.\cite{Kang2019}, as well as reducing the mass splitting. We note in particular that the overall interaction rate is highly sensitive to the value of $\delta$, with a 1\% increase (decrease) in $\delta$ producing a decrease (increase) in rate of 50\%. This is influenced by the form factor dependence on mass splitting, as well as changing the lower bound on the velocity integral, and so the exact effects of changing $\delta$ will vary model to model.\\
These new fits now potentially lie within the bounds on PICO60 given in Ref. \cite{Kang2019}, and so a comparison of the sensitivity with the different values of $\delta$ is required to understand whether or not these new models remain in reduced tension with other experimental results.\\
The use of the SHM+Stream distribution tends to lower both the cross section and mass splitting, while increasing the mass slightly, demonstrating the sensitivity of these fits to the velocity distribution in question. This motivates work with other new distributions, as well as the need to understand how this might influence detector sensitivity.

%
%

\subsection{Sensitivity}
\label{sec:sensitivity}
As discussed previously, the consideration of inelastic DM introduces a kinematic suppression of the interaction rate. As such, certain values of $\delta$, given a specific target nuclei, will greatly constrain the sensitivity of an experiment to lower DM masses. For SABRE to be considered sensitive to some combination of $m_{\chi}$, $\sigma_{0}$, and $\delta$, the signal output by the DM scattering must be significantly higher than the background reported by the experiment after applying veto, shown in Fig. \ref{fig:mc_background}. As such, observation (or lack thereof) of the modulation alone provides an easy test of the DAMA/LIBRA data. In particular, to further validate the strength of these models, both the modulating and average rate should be distinguishable from the background.

\begin{figure}[!h]
    \centering
    \includegraphics[width=0.485\textwidth]{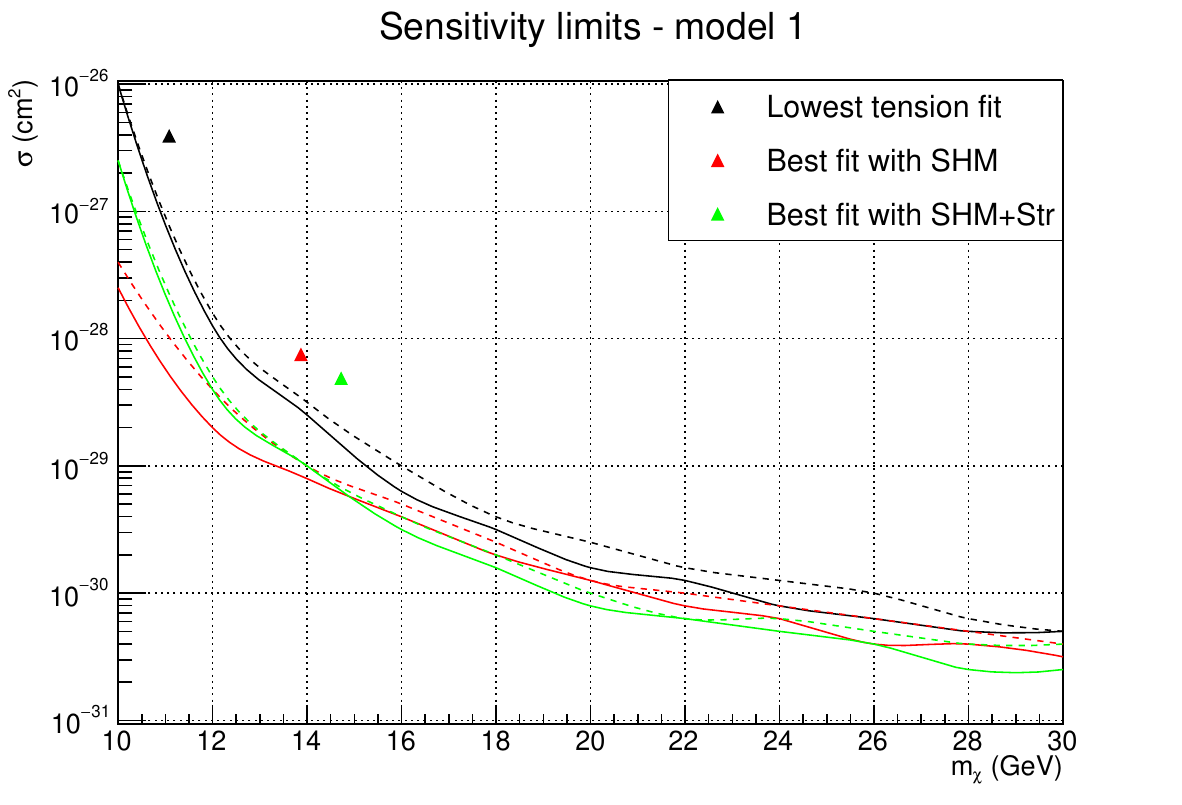}
    \includegraphics[width=0.485\textwidth]{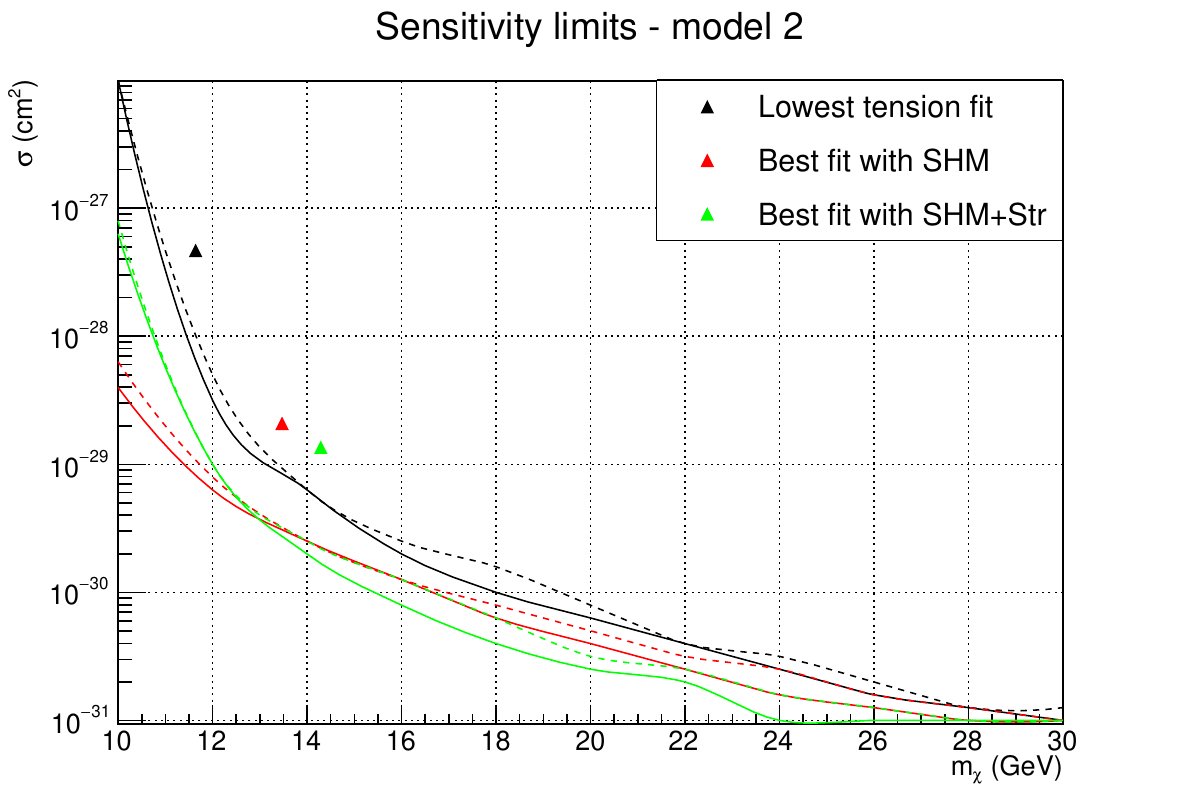}\\
    \includegraphics[width=0.485\textwidth]{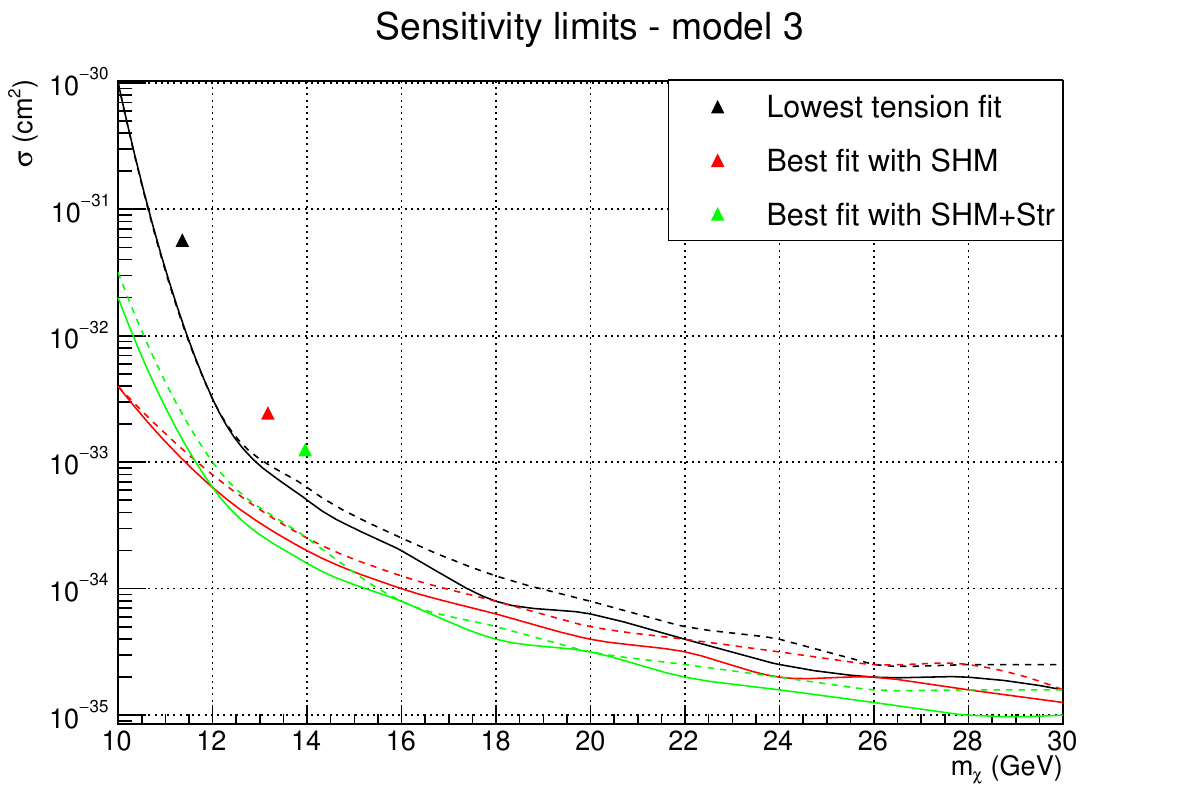}
    \caption{\textit{Left panel}: sensitivity to model 1. \textit{Right panel}: sensitivity to model 2. \textit{Bottom panel}: sensitivity to model 3. The 3 colored points indicate, each, a different fit: black for the lowest tension fit of Ref. \cite{Kang2019}, red for the best fit obtained in this work using MB speed distribution, green for our result using MB+Str speed distribution. Sensitivity of SABRE after 3 (dashed) and 5 (solid) years of data, with the different colours corresponding to values of $\delta$ from Tab. \ref{tab:proton_neutron_coupling} and MB speed distribution (black), or values of $\delta$ from Tab. \ref{tab:fit_results} and MB (red) or MB+Str (green) speed distributions. }
    \label{fig:mod1_both}
\end{figure}

If we are only interested in observing a modulating signal, the main source of modulating background will be statistical fluctuations of the average signal and total background, which could potentially create a false modulation signal that will mask the DM one. To model this, we first integrate the observed differential rate given in Eq. \ref{ob_diff} over the energy region of interest, here 1-6 keV$_{ee}$ to give the total observed rate as a function of time. This can then be projected onto $1$ and $\cos{\omega (t-t_0)}$ to separate the modulating and average components, ultimately giving an expression of the form 
\begin{equation}
    R_T = R_0+R_m\cos{\omega (t-t_0)}.
\end{equation}
The statistical fluctuations in the background are modelled by randomly sampling from a Poissonian distribution over the detector lifetime. This distribution is centred on the expected number of background counts per bin period, $N_b$, given by
\begin{equation}
    N_b = M_E \times \Delta T \times R_b \times \Delta E,
\end{equation}
where $M_E$ is the crystal mass, $\Delta T$ the width of the time bin (i.e., number of days of operation), $\Delta E$ the energy bin width, and $R_b$ the SABRE background of 0.36 cpd/kg/keV. The background modulation is then given by fitting these fluctuations to a cosine function with the same offset as DM. The sampling/fit process is then repeated 200 times and the resulting amplitudes fit to a Gaussian distribution, the mean of which, $\mu_b$ gives the background modulation. The signal+background modulation can be computed in a similar way, using a Poisson distribution centred instead on
\begin{equation}
        N_{sb}(t) = M_E \times \Delta T \times \left(\Delta E \times R_b + R_0 +R_m\cos{\omega t} \right)
\end{equation}
to construct a signal+background Gaussian distribution of fit amplitudes. Thus, the ability of SABRE to distinguish between the background only, and the signal+background modulation can be assessed by calculating the $p$-value of $\mu_b$ in the signal+background Gaussian, where for 90\% C.L, we require $p\leq0.1$. \\
Setting $\delta$ and computing $\chi^2$ for a range of $m_{\chi}$ and $\sigma_{0}$ pairs, a limit for the 90\% C.L sensitivity of SABRE can be found, taking $M_E=50$ kg, $\Delta T=30.4$ days, and $\Delta E=5$ keV for the energy range of interest here (1-6 keV). This sensitivity has been calculated for the three cases presented in Tab. \ref{couplings} for DM mass between 1 and 40 GeV, and $\sigma_0$ between $10^{-38}$ and $10^{-25}$ cm$^2$\footnote{This is much larger than the usual $\sigma_0^{SD} \sim 10^{-40}\cm^2$, $\sigma_0^{SI} \sim 10^{-45}\cm^2$ for direct detection due to the momentum suppression, that can add up a significant additional suppression, $\sim\mathcal{O}(10^{-6})$ for $v^2, q^2$ and $\sim \mathcal{O}(10^{-12})$ for $q^4, q^2v^2$.}. This is shown for the both the SHM and SHM+Str in Fig. \ref{fig:mod1_both} along with the fits to DAMA/LIBRA given in Ref. \cite{Kang2019}. These limits are given after three years of operation in dotted lines, and five in solid. 
It is clear that for all the cases considered, SABRE is well-equipped to corroborate DAMA/LIBRA's results within three to five years.\\
In all three cases we see an increase in sensitivity between three and five years of data taking, as expected. In general, there also appears to be an increase in sensitivity for both the SHM and SHM+Stream distribution, due to the reduction in mass splitting value. Interestingly, at lower masses the SHM distribution produces a signal that SABRE has greater sensitivity to compared to the SHM+Str, despite the fact that across all three models the former has a larger $\delta$. This demonstrates the influence a change in velocity distribution can have on detector sensitivity, further motivating these extensions of the analysis of the DAMA data. Additionally, it suggests that while the models presented here may be better fits to the DAMA data, they are likely to have an increased tension with other experiments, compared to those given in Ref. \cite{Kang2019}.

\section{Conclusions}
\label{sec:conclusions}

In this work we have considered proton-philic spin dependent inelastic Dark Matter models, which have been shown to reduce the tension between the DAMA/LIBRA results and other experimental collaborations. This is due to the fact that the inelastic nature of the DM constrains detectors that use low mass targets, while the proton-philic nature blinds targets like Xe and Ge.
Although it is certainly possible to carefully design models such as this one that are able to explain the lack of signal from experiments other than DAMA/LIBRA, the observed DM attributed modulation still needs to be confirmed. This can only be done with the use of a detector that utilises the same target - NaI(Tl). One upcoming detector capable of doing this is the SABRE experiment. The SABRE experiment, currently in the proof of principle stage, will have two detectors, placed in both the Northern and Southern hemispheres, and is likely to be the lowest background DD experiment with NaI target in the energy range $1$-$6\keV$. This will allow SABRE, in the case of detection of a modulation signal like in DAMA, to discriminate between a seasonal modulation, arising from yearly variation of some background, and modulation due to Dark Matter. \\
In light of this, we have computed the expected interaction rates, assuming three benchmark models, consisting of three different combinations of operators. These combinations were chosen as the ones having the lowest tension with experiments using different targets, and are mostly comprised by velocity and/or momentum suppressed operators. We have obtained and compared results using different Dark Matter velocity distributions, the usual MB distribution, and a modified speed distribution made of the combination of a MB and a stream. We have compared the resulting rates to the data from DAMA/LIBRA to find the best fit to the Dark Matter mass $m_{\chi}$, the cross section normalization, $\sigma_0$, and the mass splitting $\delta$. All models predict a best fit with $m_\chi\sim13\GeV$ and $\delta\sim20\keV$, while the normalization factor $\sigma_0$ varies depending on the model considered, due to the presence of different operators in each model. For every model, the value of $\chi^2$ per number of degrees of freedom of the fit is very low. 
Using the same rates, we have also calculated the sensitivity of the SABRE experiment, after 3 and 5 years of data taking, to all models investigated. We found that SABRE is well-equipped to detect all three models within three years. In addition to this, we note the extreme sensitivity of these inelastic models to the mass splitting parameter, where changes in $\delta$ of 1\% can increase or decrease the overall observed rate by 50\%.\\
More analysis will need to be done with these fits and the sensitivity of other detectors to determine their level of tension with other experiments. Further analysis should also be conducted with more recent measurements of the sodium quenching factor, as preliminary examinations suggest that this tends to shift the interaction peak to lower energies, likely requiring a different set of coupling constants to match the most recent DAMA data. Further studies could also be performed to see whether including analysis of the average rate $R_0$ might be better able to distinguish the fits to DAMA/LIBRA between the various models, as to date the only constraint implied is that $R_0<1$ cpd/kg/keV rather than having any clear distribution, due to the fact that DAMA has yet to release complete data for their average rate.

\begin{acknowledgments}
The authors would like to thank Andrea de Simone for his review of the paper, as well as Alan Duffy, Francesco Nuti, and Phillip Urquijo for invaluable discussions. This work was supported in part by the Australian Research Council through grants LE190100196, LE170100162, and LE160100080.  MJZ and EB are both members of the SABRE collaboration, and acknowledge the work of their colleagues in developing the Monte Carlo simulations to model detector backgrounds, the published results for which are used here.
\end{acknowledgments}

\newpage
\appendix
\section{Relevant Expressions}
\label{sec:app}

\subsection{Parameter values}
\label{sec:paramvalues}

There are a number of constants that need to be set in order to define various velocity distributions. The general terms for the SHM are based on the values used in Refs. \cite{Kang2019,shm++}. Here we define $\vec{v}_E=\vec{v}_{\odot}+\vec{v}_t$, where
\begin{equation}
\begin{split}
    \vec{v}_{\odot} &= v_{\odot}(0,0,1),\\
    \vec{v}_t &= v_t(\sin 2\pi t,\sin\gamma\cos 2\pi t, \cos \gamma\cos 2\pi t),\\
    v_{\odot} &= 232 \text{ kms}^{-1},\\
    v_t &= 30 \text{ kms}^{-1},\\
    \gamma &= \pi/3 \text{ rad}.\\
\end{split}    
\end{equation}
In this frame of reference, the DM velocity is expressed as $\vec{v} =v(\sin\theta \cos\phi, \sin\theta \sin\phi, \cos\theta)$. Assorted other constants are in Tab. \ref{tab:velocity_dists}.

\begin{table}[!h]
\centering
\begin{tabular}{@{}cccc@{}}
\toprule
\multirow{3}{*}{\textbf{SHM} \cite{Kang2019,shm++}} & DM density & $\rho$ & $0.3$ GeV cm$^{-3}$ \\  
 & Dispersion velocity & $v_0$ & $220$ km s$^{-1}$ \\
 & Escape speed & $v_{esc}$ & $550$ km s$^{-1}$ \\ \midrule
\multirow{7}{*}{\textbf{SHM+Stream} \cite{stream_dist}} 
 & DM density & $\rho$ & $0.5$ GeV cm$^{-3}$ \\
 & Escape speed & $v_{esc}$ & $520$ km s$^{-1}$ \\
 & Dispersion tensor & $\boldsymbol{\sigma}$ & $(115.3,49.9,60)$ km s$^{-1}$ \\
 & Stream velocity & $\vec{v}_s$ & $(8.6,-286.7,-67.9)$ km s$^{-1}$ \\
 & Stream density & $\rho_s$ & $0.1\rho$ \\
\bottomrule 
\end{tabular}
\caption{Values used for particular velocity distributions.\label{tab:velocity_dists}}
\end{table}

\subsection{Form Factors}
\label{sec:formfactors}

A full list of the DM form factors is available in two different forms in Ref. \cite{Liam2013,Anand2013}. The ones used in this analysis are presented here, split into their velocity dependent and independent contributions. 

\begin{table}[!ht]
\centering
\begin{tabular}{@{}ccc@{}}
\toprule
Form Factor & $F^{(ab),1}_{ij}(q)$ & $F^{(ab),2}_{ij}(q)$  \\ \midrule
$F_{4,4}^{(N,N')}$ & $C(j_{\chi})\frac{1}{16}(F^{(N,N')}_{\Sigma''}+F^{(N,N')}_{\Sigma'})$ & $0$ \\
$F_{5,5}^{(N,N')}$ & $C(j_{\chi})\frac{1}{4}\left(\frac{q^4}{m_N^4}F_{\Delta}^{(N,N')}-v_{min}^2\frac{q^2}{m_N^2}F_M^{(N,N')}\right)$ & $C(j_{\chi})\frac{q^2}{4m_N^2}F_M^{(N,N')}$ \\
$F_{6,6}^{(N,N')}$ & $C(j_{\chi})\frac{q^4}{16m_N^{4}}F_{\Sigma''}^{(N,N')}$ & $0$ \\
$F_{7,7}^{(N,N')}$ & $-\frac{1}{8}v_{min}^2F_{\Sigma'}^{(N,N')}$ & $\frac{1}{8}F_{\Sigma'}^{(N,N')}$ \\
$F_{4,5}^{(N,N')}$ & $C(j_{\chi})\frac{q^2}{4m^2_N}F_{\Sigma',\Delta'}^{(N,N')}$ & $0$ \\
$F_{4,6}^{(N,N')}$ & $C(j_{\chi})\frac{q^2}{8m_N^2}F_{\Sigma''}^{(N,N')}$ & $0$ \\ \bottomrule
\end{tabular}
\caption{Form factor contributions to differential cross sections according to Ref. \cite{Anand2013}.}
\label{tab:ff_cont_new}
\end{table}

\newpage


\label{Bibliography}
\lhead{\emph{Bibliography}} 
\setlength{\itemsep}{0pt}
\bibliography{ms} 

\end{document}